\documentclass[]{aa}
\usepackage{graphicx,epsf,psfig,epsfig}
\def\J1632{IGR~J16320$-$4751}
\usepackage{color}
\definecolor{red}{rgb}{0.7,0,0}
\definecolor{blue}{rgb}{0,0,0.7}

\begin{document}
   \title{An XMM-Newton Observation of IGR~J16320$-$4751 = AX~J1631.9$-$4752}

   \author{J. Rodriguez
          \inst{1,}\inst{2}, J.A. Tomsick\inst{3},
          L. Foschini\inst{4,}\inst{2}, R. Walter\inst{2},  A. Goldwurm\inst{1}, 
		S. Corbel\inst{5,}\inst{1}, P. Kaaret\inst{6}
}

   \offprints{J. Rodriguez : jrodriguez@cea.fr}

   \institute{CEA Saclay, DSM/DAPNIA/SAp (CNRS FRE 2591), F-91191 Gif sur Yvette Cedex, France
         \and Integral Science Data Center, Chemin d'Ecogia, 16, CH-1290 Versoix,
Switzerland
         \and Center for Astrophysics and Space Sciences, Code 0424, University of California at San Diego, La Jolla, CA 92093, USA
	\and IASF/CNR, sezione di Bologna, via Gobetti 101, 40129 Bologna, Italy 
	\and Universit\'e Paris VII, Fédération APC, 2 place Jussieu, 75251 Paris Cedex 05, France
	\and Harvard-Smithsonian Center for Astrophysics, 60 Garden
        Street, Cambridge, MA 02138, USA}

   \date{Received ; accepted}
\authorrunning{J. Rodriguez et al.}
\titlerunning{XMM observations of IGR J16320$-$4751}

   \abstract{The hard X-ray sensitivity and arcminute position 
accuracy of the
recently launched International Gamma-Ray Laboratory (INTEGRAL) has lead
to the (re-)discovery of a class of heavily absorbed hard X-ray sources 
lying in the Galactic plane.
We report on the analysis of an XMM observation of such a source 
IGR~J16320$-$4751 = AX~J1631.9$-$4752. Our analysis allowed us to obtain 
the most accurate X-ray position to date (Rodriguez et al. 2003), and 
to identify a likely infrared counterpart 
(Tomsick et al. 2003). We present the detailed analysis of the \J1632 
XMM spectra. The PN spectrum
can be well represented by a single powerlaw or a comptonized spectrum  
with a high equivalent 
absorption column density of $\sim 2\times 10^{23}$ cm$^{-2}$. 
The current analysis
and the comparison with the properties of other sources favor the
possibility that the source is a Galactic X-Ray Binary (XRB).  
The identification of two candidate IR counterparts is in good
agreement with this identification. The hard 
spectrum previously seen  with ASCA, and the brightness of the
candidate  counterparts indicate
that \J1632 is most probably a highly absorbed High Mass X-ray Binary, 
hosting a neutron star.

\keywords{Accretion, accretion disk; Stars: individual: \J1632; X-rays: binaries; X-rays: general}
   }

   \maketitle

\section{Introduction}
The International Gamma-Ray Laboratory (INTEGRAL) has been launched on 
October 17, 2002. Since then, the high sensitivity and position accuracy 
of the IBIS imager has allowed for detections 
and determinations of arcminute positions for faint hard X-ray sources 
(e.g. IGR J16318$-$4848, Courvoisier et al. 2003; IGR J16358$-$4726, 
Revnivtsev et al. 2003, in addition to \J1632). 
It is interesting to note that these sources belong  
to a class of highly absorbed low luminosity X-ray sources, which renders 
their detection in the soft X-rays ($\leq 10$ keV) difficult. 
In fact, some of them  
were missed with the All Sky Monitors of  past X-ray missions, especially 
those sensitive in the 
1$-$10 keV spectral range. As a result, these sources have remained  
poorly studied until the advent of INTEGRAL, and there may be many more than 
previously realized. In that view the 
IBIS/ISGRI detector on board INTEGRAL appears perfectly suited since it 
works in a spectral 
range ($\geq 15$ keV) not affected by absorption. Once such a source is 
detected, 
given the good position accuracy of the IBIS/ISGRI detector, it is possible 
to use 
highly sensitive soft (1-10 keV) X-ray telescopes, such as XMM-Newton or 
CHANDRA,
to 1) obtain a more precise position and allow for counterpart search, 
and 2) obtain a 
soft X-ray spectrum and try to identify the type of the source. 
Such studies should then 
allow for a better understanding of the nature of these highly
absorbed sources and the physics underlying the emission/absorption processes 
(e.g. Revnivtsev et al. 2003b). \\   
\indent \J1632 was detected on Feb. 1.4 UT (Tomsick et al. 2003a), as 
a hard X-ray 
source with the IBIS/ISGRI detector (Lebrun et al. 2001) on board INTEGRAL
at R.A.$_{J2000}$=16$^\mathrm{h}$32$^\mathrm{m}$.0, 
$\delta_{J2000}$=-47$^{\circ}$51$^\prime$ ($\pm$2$^\prime$), during an 
observation of the Galactic 
Black Hole Candidate (BHC) 4U 1630$-$47 (PI Tomsick). The source was observed 
to vary 
significantly in the 15$-$40 keV energy range on time-scale of $\sim 1000$ s,
and was detected in some occasions above 60 keV (Tomsick et al. 2003a). 
This source has a position consistent with that of AX J1631.9-4752, which
was observed with ASCA in 1994, and 1997. The ASCA spectrum was fitted with 
a powerlaw with a hard photon index (0.2 $\pm0.2$, Sugizaki et al. 2001), 
which 
may suggest that the source belongs to the High Mass X-ray Binary (HMXB) 
class. Analysis
of archival BeppoSAX-WFC data revealed that this source was 
persistent for at
least 8 years (in't Zand et al. 2003). Their $2-28$ keV spectral 
analysis shows a quite 
different result, since they obtain a soft photon index (2.5 $\pm 0.3$) for 
the powerlaw. 
This evolution and the persistence of the source may indicate the presence
of an absorbed XRB. Two possible infrared counterparts  have been identified 
(Tomsick et al, 2003b).\\
\indent We report here the detailed spectral analysis of an XMM public Target 
of
Opportunity (ToO) observation of \J1632, and compare it to the former 
observations 
of the source. In Sec. \ref{sec:xmmdatared} we provide
details about the XMM observation and data reduction methods that were 
employed for the analysis.
We describe the spectral analysis in Sec. 3, and  present results on the time variability of 
the source in Sec. 4. The infrared counterparts will be discussed in Sec. 5, and the results 
of our analysis will be discussed in the last section of this paper.
\section{XMM data reduction and analysis}
\label{sec:xmmdatared}
\J1632 was observed with XMM-Newton on March 4, during a public ToO
pointing on the INTEGRAL position that started around 21 h UTC. 
The data were processed using the Science Analysis Software v.5.4.1. 
Images were then obtained both from the EPIC
MOS (Turner et al. 2001) and EPIC PN (Str\"uder et al. 2001) cameras.
The EPIC-PN was operating in imaging mode with large window and 
a medium filter, the
EPIC MOS2 in imaging mode with a full window and a medium filter. The
EPIC-MOS1 was operating in timing mode (medium filter).  
During the processing, the data were screened by rejecting periods of 
high background, and by filtering the bad pixels. 
Correction for vignetting (Lumb 2002) has not been applied, because 
the source is close to 
the center of the field of view ($< 2\arcmin$). 
One source was detected by the MOS2 within the 1 arcmin INTEGRAL error circle. 
The source position (obtained  following the procedures given in the 
Introduction to XMM-Newton Data Analysis\footnote {Snowden et al. 
http://xmm.vilspa.esa.es/external/\\xmm\_sw\_cal/sas\_frame.shtml})  
 is $\alpha_{J2000}$=16$^\mathrm{h}$ 32$^m$ 01.9$^\mathrm{s}$ and $\delta_{J2000}$=-47$^{\circ}$
52$^{\prime}$ 29$^{\prime\prime}$ ($\pm$4$^{\prime\prime}$ at the 
90$\%$ confidence level, Rodriguez et al. 2003). 
It should be noted that 
this position is also consistent with the ASCA (at 0.4$^{\prime}$ from the
XMM position with an uncertainty of 1$^{\prime}$), and BeppoSAX 
(at 0.7$^{\prime}$ from the XMM position, uncertainty of 1.7$^{\prime}$)
positions of AX J1631.9$-$4752.\\
\indent Due to soft proton flares during this observation, the MOS2 data
 are not usable for spectral and timing analysis, 
and only $\sim$ 4.9 ks out of a total of 25 ks are exploitable for
scientific (spectral and timing) studies with the PN camera. 
The spectrum and light curve for the source were extracted from a 
circular region centered on 
the source with a radius of 45$^{\prime\prime}$ (which 
gives an encircled energy
fraction of about 85 $\%$). The background spectrum and light 
curve were extracted
from a source free region, with a radius of 2 arcmin.
The response matrices were generated with the SAS package
(arfgen, rmfgen). The spectrum was grouped with a minimum of 25 counts
per channels and was fitted with the XSPEC v11.2 package.

\begin{figure}[htbp]
\centering
\epsfig{file=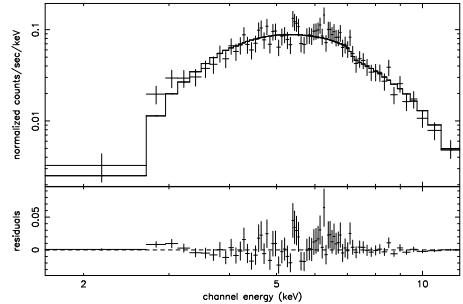,width=6cm, height=4cm}
\caption{ EPIC-PN spectrum of \J1632 and residuals to the absorbed powerlaw model. }
\label{fig:spec}
\end{figure}

\section{Spectral studies}
\begin{figure}[htbp]
\epsfig{file=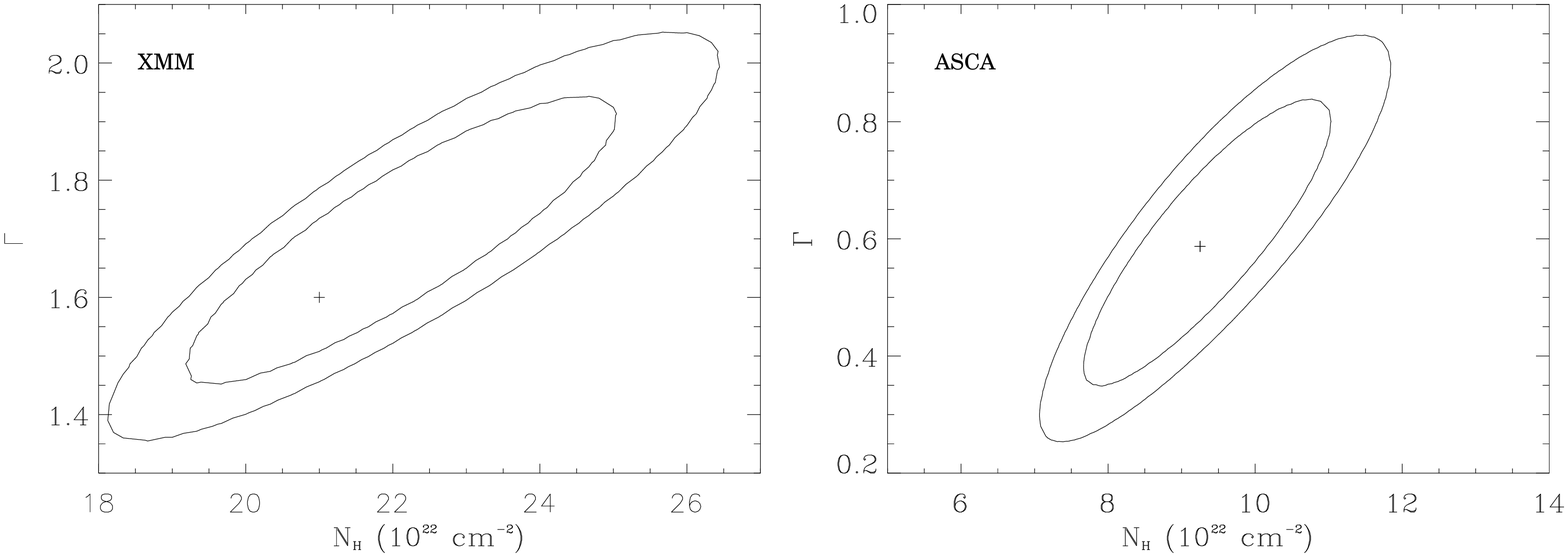,width=\columnwidth,height=4cm}
\caption{{\bf{Left:}} Error contours for the column density (N$_\mathrm{H}$), and the 
photon index ($\Gamma$) obtained 
from the XMM-Newton observation. 
The cross marks the location of the best fit values, and the 68$\%$ ($\Delta\chi^2$=2.30) and 90$\%$ 
($\Delta\chi^2$=4.61) contours are shown. {\bf {Right: }} Same plot for the ASCA observation.}
\label{fig:conf}
\end{figure}

The PN spectrum was fitted with different models.
In a first run, only single component models were tested.
 An absorbed powerlaw gives a good representation of the spectrum,
 with a reduced $\chi^2$ of 0.81 (Rodriguez et al. 2003 \& Fig. \ref{fig:spec}).
An absorbed blackbody or disc blackbody give acceptable fits,
with a reduced $\chi^2$ of 0.81 (bbody) and  0.83 (diskbb), but the temperature
returned from the fits (kT=2.1 keV for the bbody and 3.8 keV for the diskbb) 
is a bit higher (especially for the diskbb component) than what is observed in general 
for accreting compact objects, even during the
high luminosity states where a thermal component contributes strongly
to the spectrum (see Tanaka \& Shibazaki 1996 for a review).
Furthermore, the detection of the source at higher energies ($\geq 15$ keV, 
Tomsick et al. 2003a) indicates the need for an additional component to account
for this hard part. A comptonized spectrum
(comptt, Titarchuk 1994) gives a good representation of the spectrum,
with a reduced $\chi^2$ of 0.83. The spectral parameters are not well
constrained, however, if they are all left free. Fixing the electron
temperature to 10 keV (which is an reasonable value, Barret 2001),
leads to a reduced $\chi^2$ of 0.84, with spectral parameters
compatible with what is commonly observed in accreting neutron stars (Table
\ref{tab:param}, \& Barret 2001). Since both models are strongly correlated to
N$_\mathrm{H}$, the 68$\%$ and 90$\%$ confidence intervals are shown on Fig. \ref{fig:conf},
for the photon index and the absorption column density.\\
\indent We have re-analysed the publicly available ASCA data, and fitted the SIS and GIS energy
spectra simultaneously in XSPEC, with a simple model of an absorbed powerlaw.
Our best result gives an equivalent absorption column density of
N$_\mathrm{H}=(9.2\pm 1.1) \times 10^{22}$ cm$^{-2}$,
 and a photon index of $\sim 0.6 \pm0.2$ (errors at 1$\sigma$), 
somewhat softer than that obtained by Sugizaki et al (2001). The error contour
 plot is shown on Fig. \ref{fig:conf}, allowing for a direct comparison with 
that of the XMM observation. The BeppoSAX spectrum was fitted with an 
absorbed  powerlaw of photon 
index 2.5, and $N_H=20\times10^{22}$ cm$^{-2}$ (in't Zand et al. 2003).
\begin{table}[htbp]
\caption{Spectral fit parameters. The given fluxes are corrected for
  an encircled energy fraction of 85$\%$. The errors are the 1$\sigma$ 
confidence level.}
\begin{tabular}{c c}
\hline
\hline
\multicolumn{2}{c}{Power-law Model}\\
\hline
N$_\mathrm{H}$ & 21$_{-1}^{+4}\times 10^{22}$ cm$^{-2}$\\
$\Gamma$ & $1.6_{-0.1}^{+0.2}$\\
$\chi^2$ & 55.9 (69 dof)\\
2$-$10 keV unabsorbed flux& $1.7\times 10^{-11}$ erg/s/cm$^2$\\
\hline
\multicolumn{2}{c}{Comptt}\\
\hline
N$_\mathrm{H}$ & 14$\pm2\times 10^{22}$ cm$^{-2}$\\
kT$_{\mathrm{seed}}$ & $1.6_{-0.3}^{+0.5}$ keV\\
kT$_{\mathrm{e}}$ & 10 keV fixed\\
$\tau$ & $\leq 3.35$ (3 $\sigma$)\\
$\chi^2$ & 56.2 (68 dof)\\
2$-$10 keV unabsorbed flux&$1.1 \times 10^{-11}$ erg/s/cm$^2$\\
\hline
\end{tabular}
\label{tab:param}
\end{table}

\section{Timing behavior}
\begin{figure}[htbp]
\centering
\epsfig{file=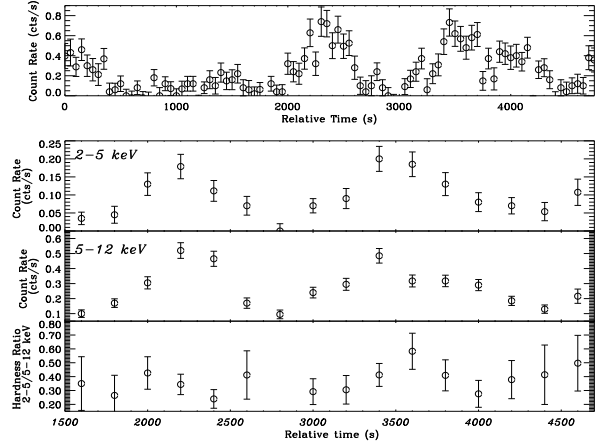,width=\columnwidth,height=6cm}
\caption{Upper Panel: 0.5-12 keV PN light curve of \J1632 measured
  after background subtraction.  The time sampling
is 50s, which shows the fast rise during the flare episodes. The lower
panels represent a zoom on the region of the flares (starting around
relative time 1500s). The 2-5 keV, 5-12 keV and hardness ratio between
2-5 and 5-12 keV are shown. The time sampling is 200 s.}
\label{fig:lite}
\end{figure}

The PN light curve is displayed in Fig. \ref{fig:lite}. The source
shows variability on a time scale of $\sim 50-100$ s. A slow decrease 
and a period of relatively quiet emission is first observed until 
$\sim$ 2000 s.
Then, the source undergoes two flares: the first starts at 2000 s after time 0 
(begining of the good time intervals),  
reaches its maximum about 300 s later. The flux increases by  
a factor of 2.3 in the mean time. This flare lasts for
$\sim 900$ s. The second flare starts around 3050-3100 s, 
and reaches the maximum around 3450 s, the flux increases by a 
factor of $\sim3.3$, during this time. 
A third flare may start around 4500 s, but our observation stops soon after, 
and does not allow us to follow it.\\
\indent We produced light curves in two different energy bands (2$-$5 keV
and 5$-$12 keV, Fig.\ref{fig:lite}), and the hardness ratio between
these two. The flare episodes occur in the two ranges in a similar
manner (Fig. \ref{fig:lite}). With the time binning of 200s, the 
flux increases by a factor of $\sim4$ 
between 2 and 5 keV, and $\sim3$ between 5 and 12 keV during a time 
interval of 400 s 
(first flare), and by  a factor of $\sim 3$ between  2 and 5 keV, and $\sim2$ between 5 and 12 keV 
during 400 s (second flare)\footnote{Note that the differences of the variation rates compared 
 to those 
of the total range light curve, are simply due to the different time binnings used for the 
light curves.} 
This similarity indicates that the flares 
are related to broadband flux increase rather than
variations of the absorption (since the hard band would be much less
affected in this case). The hardness ratio does not show significant
changes between the low flux periods and the flares. The same behavior is 
observed if
the light curves are produced in other energy bands (e.g. 2$-$3 keV and 
2$-$4 keV).\\
\indent We searched for pulsations and quasi-periodic oscillations in the
power spectra of the source. However, the low counting statistics and
short exposure time does not enable us to obtain strong constraints. 
Indeed taking into account a net average counting rate of
0.22 cts/s for the source, and 0.16 cts/s for the background, during a
4.8 ks observation, leads to a 3$\sigma$ upper
limit on the amplitude of a periodic signal between $\sim 5$ mHz and 10 Hz
(where the power spectrum is dominated by Poissonian noise)  
 of 12.25 $\%$. The 3$\sigma$ upper limit for QPOs is higher than this 
value in the given frequency range (due to the non-zero width of the QPO).

\section{Infrared Counterpart}
\begin{figure}[htpb]
\centering
\epsfig{file=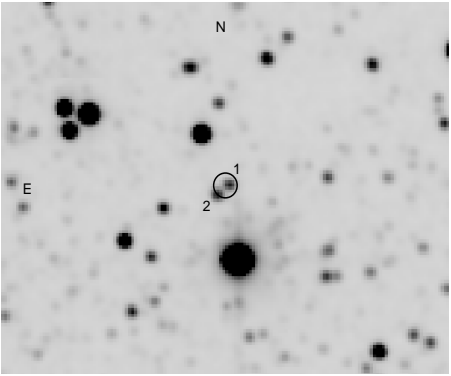,width=6cm,height=5cm}
\caption{2-MASS K-band $2.5\times 2$$^{\prime}$ image of \J1632 field, the XMM error circle is
  superimposed.  One can clearly see the two possible counterparts. 1 stands for the most probable 
one, and 2 for the second.}
\label{fig:counter}
\end{figure}

From the improved position obtained with XMM$-$Newton (Rodriguez et al. 2003), two candidate 
infrared counterparts have been identified (Tomsick et al. 2003b, Fig.~\ref{fig:counter}). 
The first one is located at $\alpha_{J2000}$ = 16$^\mathrm{h}$ 32$^\mathrm{m}$ 01$^\mathrm{s}$.75, 
$\delta_{J2000}$ = -47$^{\circ}$
52$^{\prime}$ 28$^{\prime\prime}$.9 (1$^{\prime\prime}$ uncertainty) with magnitudes\footnote{
the magnitudes come from the ``2MASS All-Sky
Point Source Catalog'' http://www.ipac.caltech.edu/2mass/ } 
$K_s = 10.99\pm 0.03$, $H = 13.03 \pm 0.04$, 
$J > 14.08$ (97 $\%$ confidence). 
The second source ($K_s = 10.82\pm0.04$, $H = 11.24\pm0.03$, $J = 12.13\pm0.02$) 
is on the southeast 
edge of the X-ray error circle. From the Palomar Observatory Sky Survey (POSS) 
epoch I and II, we can derive $R>21$ for the first source,
and R=14.6 ($\pm 0.3$, 1$\sigma$ error POSS epoch I, period 1949-1965) and R=15.4 
($\pm 0.3$, $1\sigma$ error POSS epoch II, period 1985-2000), for the second one (note 
that the evidence for variability is only marginal). 
To estimate the equivalent absorption column density along the line of
sight we used a web-based tool that uses data from  Dickey \& Lockman (1990).
For the position of \J1632, the average value is $\sim 2.1\times
10^{22}$ cm$^{-2}$. Then, 
assuming A$_\mathrm{V}=5.6\times10^{-22}\times $N$_\mathrm{H}$ (Predehl \& Schmitt 1995), we derived
A$_\mathrm{V}=11.7$. With this value, we can calculate the 
dereddened fluxes in the three bands, and compare them to tabulated objects. 
The second source appears to be a star with a peak in the $J$ band. The colors $(J-H)=0.9$ and
$(H-K)=0.4$ are typical for a M$-$type star with a temperature around 3000~K.\\
\indent With this visual extinction, the first source, the faintest and most probable counterpart, 
presents some very interesting features. The dereddened flux increases toward radio wavelenghts 
(from $J$ to $K$ band)
and the colors $(J-H)>3.1$ and $(H-K)=2.0$ suggest an infrared excess that is likely to be
due to circumstellar matter (probably hot plasma or warm dust). However, the absorption
column density along the line of sight may be different from this value. The N$_\mathrm{H}$ 
returned from 
XMM spectral fits leads to an extinction of A$_\mathrm{V}=106$, giving a dereddened $K_s$ 
magnitude close
to 0 for both objects. In addition, some molecular clouds and inhomogeneities in the ISM can 
make the A$_\mathrm{V}$ increase. If, for example, we take a value of  
A$_\mathrm{V}$=30 (i.e N$_\mathrm{H}=5 \times10^{22}$ cm$^{-2}$), the dereddened magnitudes 
would suggest a K-giant or 
supergiant, rendering the object more usual. Furthermore, the wavelength dependence of the
interstellar extinction is poorly constrained when studying sources close to the Galactic ridge. We 
thus cannot exclude other types for the companion star. We remark for example that the color, and 
magnitude of the second candidate counterpart could indicate a high mass star (O or B spectral type).

\section{Discussion}
From their analysis of ASCA data, Sugizaki et al (2001) suggested that
AX~J1631.9$-$4752 was an HMXB with a pulsar, given the flat spectrum they
obtained.  Our best model, a powerlaw or a comptt spectrum, is also
consistent with the source being  a Galactic XRB.
A comptonized spectrum would not be surprising for such an object, but alone 
it is not a sufficient 
argument since extragalactic sources have similar X-ray spectra. For
a distance range of 5$-$15 kpc (the latter giving an upper limit on the 
distance for the object to be Galactic), the powerlaw model leads to  
unabsorbed 2$-$10 keV isotropic luminosities of 
$0.5$$-$$4.5\times 10^{35}$ erg/s, which may be compatible with the observe
 luminosity of either a 
neutron star or a black hole in a low hard state. The spectral parameters we 
obtain are also compatible with those commonly observed for this kind of 
objects
(e.g. Tanaka \& Shibazaki 1996). The persistence of the source over 8 years
 (in't Zand et al. 2003), and the evolution of the spectral parameters 
resembles the spectral transitions seen in XRBs.\\
\indent It is interesting to note the
similarities of \J1632 with the
persistent X-ray source 1E 1743.1$-$2843 whose primary type is still a 
matter of debate. Both sources are persistent, heavily absorbed, and
undergo XRB-like spectral transitions between hard states 
(Porquet et al. 2003), and softer ones 
(Cremonesi et al. 1999). Although for 1E 1743.1$-$2843 the neutron star
 hypothesis could 
not be ruled out, Porquet et al. (2003) have suggested that this object
 might be a black hole in
a hard state. In the case of \J1632, the BeppoSAX and XMM observations alone 
would rather indicate a persistent black hole undergoing spectral 
transitions between 
hard states and softer ones. However, the low value of the photon
index during the ASCA observation,
and its large variations between ASCA (Sugizaki et al 2001), 
BeppoSAX (in't Zand et al 2003), and XMM (current study) observations 
are rather 
unusual for a black hole. On the other hand, such very hard photon indices 
and large 
variations (from 0 to 2.4) have already been observed in the neutron star 
system 
Sco X$-$1 (D'Amico et al 2001), but also in some other sources (Sugizaki et al.
 2001). 
In addition, the powerlaw model gives extrapolated 1$-$20 keV
luminosities (between 5$-$15 kpc), of 0.93$-$8.4 $\times 10^{35}$ erg/s, and  
extrapolated 20$-$200 keV
luminosities of 0.23$-$2.1 $\times 10^{36}$ erg/s (the comptt model leads to
slightly lower values). With these values, the
source would lie in the neutron star box on Fig. 1 of Barret et al. (1996).
The relatively good fit obtained with the 
comptt model is compatible with a Comptonization of soft photons on hot 
electrons surrounding 
the compact object. However, the lack of data above 12 keV, 
does not allow us to obtain a good constraint on the cut-off energy.  
The detection of hard  emission with INTEGRAL (Tomsick et al. 2003a) 
 appears in good agreement with a comptonized spectrum. The hard photon index obtained 
during the ASCA observation, however, is not easy to understand in the framework of thermal 
Comptonization. \\
\indent The identification of the infrared counterpart remains
difficult given the proximity of the two candidates. The faintest one 
(number 1 on Fig. \ref{fig:counter}), however, is more likely related to 
the X-ray source. 
This candidate could either favor a reprocessing of high energy photons 
by a cloud or some dust, or the emission
from a supergiant or K-giant star depending on the visual extinction on the 
line of sight. 
Both hypothesis are in good agreement with the high absorption obtained 
from the X-ray
spectrum, and the spectral parameters obtained from our fits, which are 
similar to those 
observed in the case of HMXB (e.g. GX 301$-$2, Saraswat et al. 1996, 4U1700$-$37,
 Boroson et al. 2003),
although we do not detect lines in \J1632 with a 3$\sigma$ upper 
limit on the equivalent width of a narrow 6.4 keV Iron (emission) line 
of 112 eV.
The position of \J1632 in the Galactic plane, and along a spiral
 arm (the Norma arm),  a region of star formation, where a number of young (massive) 
stars can be found, appear in good agreement with the system being an HMXB. 
 It is interesting to note the change of equivalent absorption density between 
the different observations, (from $\sim $9$\times 10^{22}$ cm$^{-2}$ to 
$\sim $20$\times 10^{22}$ cm$^{-2}$). This strongly favors an absorption 
intrinsic to the source,
 similar to what observed in heavily absorbed sources (e.g. Revnivtsev et al. 2003 
and reference therein). 
In the case of a Galactic XRB, the absorption could be due to the
accretion flow, and the changes could be associated with variations of the accretion rate. \\
\indent Although the source appears more likely to be a Galactic XRB (given 
its similarities with known neutron stars or black holes XRB) we can not exclude
totally an extragalactic source, e.g. an ULX seen through the Galactic plane.
Assuming a typical luminosity of $10^{40}$ erg/s (Foschini et al 2002), the simple 
power law model would lead to a distance to the source of about 2.2 Mpc.
Although compatible with an extragalactic origin, this hypothesis appears 
not to be favored given such a short distance.  Longer X-ray observations 
and more multiwavelength studies (e.g. infrared spectra), should allow for more
precise answers to all the above mentioned points. Note that a better X-ray position
would be useful to select a unique IR counterpart, allowing for a better understanding of 
the nature of the source.

\begin{acknowledgements}
J.R. thanks D. Barret, S. Chaty, Y. Fuchs, A. Paizis, and M. Revnivtsev,  for useful discussions 
and careful reading of the manuscript. The authors thanks the anonymous referee for useful
 comments which allowed to improve the quality of the paper. J.R. acknowledges financial support 
from the French 
spatial agency (CNES). J.A.T. acknowledges partial support from NASA grant NAG5-12703. 
This work is based on observations obtained with XMM-Newton, an ESA 
science mission with 
instruments and contributions directly funded by ESA Member States and the USA (NASA). 
 This publication makes use of data products 
from the Two Micron All Sky Survey, which is a joint project of the University of 
Massachusetts and the Infrared Processing and Analysis Center/California 
Institute of Technology, funded by the NASA and the National Science
Foundation. This reasearch has made use of The Digitized Sky Surveys 
that were produced at the Space Telescope Science Institute under U.S. 
Government grant NAG W-2166.

\end{acknowledgements}

\end{document}